\documentstyle{article}\begin{document}
\title{Relativistic diffusion of massless particles }
\author{Z. Haba \\
Institute of Theoretical Physics, University of Wroclaw,\\ 50-204
Wroclaw, Plac Maxa Borna 9, Poland,
\\email:zhab@ift.uni.wroc.pl}\date{\today}\maketitle
\begin{abstract}
We obtain a  limit when mass tends to zero of the   relativistic
diffusion  of Schay and Dudley. The diffusion process has the
log-normal distribution. We discuss Langevin stochastic
differential equations leading to an equilibrium distribution. We
show that for the J\"uttner equilibrium distribution the
relativistic diffusion is a linear approximation to the Kompaneetz
equation describing a photon diffusion in an electron gas.
 The
stochastic equation corresponding to the J\"uttner  distribution
is explicitly soluble. We relate the relativistic diffusion to
imaginary time quantum mechanics. Some astrophysical applications
(including the Sunyaev-Zeldovich effect) are briefly discussed.
\end{abstract}
 \section{Introduction}
 An extension of the idea of diffusion to relativistic
 theories poses some problems of conceptual as well as
 physical character (for a discussion of these problems and different
 approaches to solve  them see
 (\cite{schay}\cite{dudley}\cite{hakim}\cite{deb}\cite{dunkel}\cite{hor1}; for
a review and further references see \cite{talkner}\cite{deb2}). In
\cite{haba} we developed the diffusion  theory of Schay
\cite{schay} and Dudley\cite{dudley} of massive relativistic
particles discussing  friction terms leading to an equilibrium. In
refs.\cite{schay}\cite{dudley} the relativistic Brownian motion is
uniquely defined by the requirement that this is the diffusion
whose four momentum stays on the mass-shell. The formulas become
singular as $m\rightarrow 0$.

 The photon is a massless particle which could experience
 a diffusive behavior.
 We could think of a multiple scattering of the photon in a charged gas as
 resembling the Brownian motion. The number of such encounters must be large
  if the approximation is to be feasible.
 This happens in astrophysical applications when the photon travels over large distances
 in intergalactic space filled by a ionized gas.
 The diffusive approximation could also apply to the early stages of the Big Bang
  and to the photon motion inside the plasma shell of the compact stellar objects (stars, black holes)
The relativistic Boltzmann equation  is the standard tool in a
study of the evolution of the photon in a gas of charged particles
\cite{rybicki}\cite{dodelson}. The diffusion approximation to the
Boltzmann equation has been applied in many models of the
transport phenomena \cite{landau}. In the context of the photon
propagation such a diffusion approximation is known as the
(non-linear) Kompaneetz equation
\cite{komp}\cite{wymann}\cite{rybicki}. The Kompaneetz equation
arises as an approximation to the Boltzmann equation describing an
evolution of the photon  in a gas of electrons. In such a case the
evolution of a beam of photons is disturbed by Compton scattering
and Bremsstrahlung. These are the processes widely studied in
astrophysics \cite{rybicki}\cite{ens}\cite{itoh}.

In this paper we investigate a diffusion of massless particles
without spin in the framework of classical relativistic diffusion
theory . We show that there is a massless counterpart of Schay
\cite{schay} and Dudley \cite{dudley} relativistic diffusion (the
formula for a diffusion of a massless particle has been also
derived in \cite{sorkin} as a continuum limit of   discrete
dynamics of space-time, see also \cite{mex}). We derive the
diffusion as a classical limit of quantum mechanics (the Boltzmann
equation usually also incorporates quantum scattering processes).
Such a diffusion process has no limit as time grows to infinity.
We discuss (sec.3) drags which lead to an equilibration of the
process. Subsequently, we show in sec.4 that if the Kompaneetz
equation is expanded around its equilibrium solution (J\"uttner or
Bose-Einstein) and non-linear correction terms are neglected then
the resulting diffusion equation is the same as our relativistic
diffusion with the friction leading to the same equilibrium. In
secs.5 and 6 we discuss the dynamics in an affine time parameter.
It is shown that the dynamics (in particular, the approach to the
equilibrium) can be well controlled by methods of Langevin theory
(sec.5) or quantum mechanics (sec.6). In the final sec.7 we
discuss the problem of time in the relativistic diffusion. We show
that it can be approached in a similar way as in the deterministic
relativistic dynamics. The affine parameter on the trajectory can
be replaced by the (random) laboratory time. As a result from a
solution of the diffusion equation in the affine time we can
obtain a solution of the transport equation involving only the
physical variables (in particular, the laboratory time).
\section{Quantum origin of the relativistic diffusion equation} In
our earlier paper \cite{haba}, following Schay\cite{schay} and
Dudley\cite{dudley}, we have defined the generator of the
relativistic diffusion as the second order differential operator
on the mass-shell ${\cal H}_{m}$
\begin{equation}
p^{2}=p_{0}(\tau)^{2}-p_{1}(\tau)^{2}-p_{2}(\tau)^{2}-p_{3}(\tau)^{2}=m^{2}c^{2}
\end{equation}
The generator is singular at $m=0$. For this reason in order to
define a diffusion of massless particles we propose another
method. A diffusion should be considered as an approximation of a
complex multi-particle dynamics. In our case the relativistic
dynamics. It is not clear how to describe many particle systems in
a relativistic way. Quantum field theory leads to a relativistic
description of scattering processes. However, its reduction to
particle dynamics has not been explored. The photon as a massless
particle is a typically relativistic object. At the same time this
 is a quantum system with an internal angular momentum (spin). We
should describe it by a relativistic wave function. In this paper
we neglect the spin. In \cite{habajpa} we discuss the diffusion of
particles with a spin (the helicity in the massless case). It is shown in \cite{habajpa}
that the dissipative part of the evolution  does not depend on the helicity
if $m=0$. Hence,
the neglect of spin in this paper is justified.

The explicitly Lorentz invariant description of dynamics must in
fact be static. It has to be described in terms of space-time
trajectories. The laboratory time $x^{0}$ is one of the
coordinates. Hence, the evolution as a function of this coordinate
should be inferred at the later stage from a study of the set of
paths. In the massive case the position of a particle on the
space-time path is most conveniently described by its own time :
the proper time. If a particle is massless then the notion of the
proper time as a time of an observer moving with the particle does
not make sense. Nevertheless, it is still convenient to introduce
an affine parameter $\tau$ describing the fictitious motion along
the particle trajectory (we shall still call it a proper time
although no observer can move together with the particle).

The proper time as an additional parameter in a description of a
relativistic quantum particle has been introduced in
\cite{stuck}\cite{feyn} (in quantum theory the relation to the
time of an observer moving with the particle is  lost anyhow). It
can be treated as a convenient tool in a formulation of
relativistic quantum mechanics and  quantum field theory. The
equation for a massless free field is
\begin{equation}
i\partial_{\tau}\phi=(\partial_{0}^{2}-\triangle) \phi
\end{equation}
where $\partial_{0}^{2}-\triangle$ is the wave operator.

 We may consider a
perturbation of (2) by some other fields (an environment) so that
eq.(2) remains invariant under the Lorentz group. When we average
over the environment then in the Markovian approximation  the
system  will be described by the master equation for the density
matrix $\rho$. A preservation of the trace and positivity of the
density matrix requires that the master equation should have the
Lindblad form \cite{lind}. Then, the Lorentz invariance and an
assumption that the Lindblad generators are built from the Lorentz
generators leads in the simplest (linear) case to the  equation
\begin{equation}
i\partial_{\tau}\rho=[\partial_{0}^{2}-\triangle,\rho]+\frac{1}{4}\gamma^{2}[M_{\mu\nu},[M^{\mu\nu},\rho]]
\end{equation}
where $M_{\mu\nu}$ are the generators of the algebra of the
Lorentz group. Eq.(3) can be rewritten as an equation
  for the Wigner function \cite{zachar} (proper time equations for the Wigner
  function in relativistic quantum mechanics appear in \cite{hor2} and in quantum field theory  in
  \cite{kelly}\cite{elze})
\begin{equation}
\partial_{\tau}W=p^{\mu}\frac{\partial}{\partial
x^{\mu}}W+\frac{1}{4}\gamma^{2}L_{\mu\nu}L^{\mu\nu}W
\end{equation}
where
\begin{equation}
L_{\mu\nu}=-i(p_{\mu}\frac{\partial}{\partial
p^{\nu}}-p_{\nu}\frac{\partial}{\partial p^{\mu}})
\end{equation}
is a realization of the algebra of the Lorentz group in the
momentum space.

 It
can be checked that the diffusion generator of \cite{haba} can be
expressed as
\begin{equation}
\triangle_{H} =\frac{1}{2}L_{\mu\nu}L^{\mu\nu}
\end{equation}
The derivation of the relativistic diffusion equation from quantum
mechanics is similar  to a derivation of the Boltzmann equation
from quantum mechanics  in \cite{erdos}. Eq.(4) makes sense in the
massless case. Choosing ${\bf p}$ as coordinates on ${\cal H}_{0}$
we obtain
\begin{equation}
\triangle_{H}=p_{j}p_{k}\partial^{j}\partial^{k}
+3p^{k}\partial_{k}
\end{equation}
where $k=1,2,3$ and $\partial^{j}=\frac{\partial}{\partial
p_{j}}$. We note that the generator (7) is the limit $m\rightarrow
0$ of $m^{2}c^{2}\triangle_{H}$ of ref.\cite{haba}.

 The relativistic diffusion
equation in the momentum space (a relativistic analog of the
Brownian motion) reads
\begin{equation}
\partial_{\tau}\phi_{\tau}=\frac{1}{2}\gamma^{2}\triangle_{H}\phi_{\tau}
\end{equation}
$\gamma^{2}$ is a diffusion constant which has the dimension of
$\tau^{-1}$ (our choice in the next section:momentum divided by
length).

\section{An approach to the equilibrium }
 We add a drag term
\begin{equation}
X=Y_{j}\frac{\partial}{\partial
p_{j}}+\gamma^{2}R_{j}\frac{\partial}{\partial
p_{j}}+p_{\mu}\frac{\partial}{\partial x_{\mu}}
\end{equation}
to the diffusion (8). The vector field (9) defines a perturbation
of the relativistic dynamical system (at $\gamma=0$)
\begin{equation}
\frac{dp_{j}}{d\tau}=Y_{j} \end{equation} \begin{equation}
\frac{dx^{\mu}}{d\tau}=p^{\mu} \end{equation} where from eq.(1)
(for $m=0$) we have $p^{0}=\vert {\bf p}\vert$. The zero component
of eq.(11) determines a relation between the parameter $\tau$ on
the trajectory and the laboratory time $x^{0}$. Let
\begin{equation}
{\cal G}=\frac{1}{2}\gamma^{2}\triangle_{H}+X
\end{equation}
A solution of the diffusion equation
\begin{equation}
\partial_{\tau}\phi_{\tau}={\cal G}\phi_{\tau}\end{equation}
with the initial condition $\phi(x,{\bf p})$ defines a
distribution of random paths (see sec.5) starting from the point
$(x,{\bf p})$ of the phase space. At $\gamma=0$ we have a
distribution of deterministic trajectories. The dynamics in the
laboratory time $x^{0}$ must be derived from the  static picture
of space-time paths ( see the discussion of relativistic
statistical dynamics in \cite{hakim}).

 We are interested in the behavior of the relativistic dynamics at large
 values of $\tau$. In particular, whether the diffusion generated by
${\cal G}$ can have an equilibrium limit. We say that  the
probability distribution $dxd{\bf p}\Phi$ is the invariant measure
for the diffusion process (see \cite{ikeda}) if \begin{equation}
\int dxd{\bf p}\Phi(x,{\bf p})\phi_{\tau}({\bf p},x)\equiv\int
dxd{\bf p}\Phi_{\tau}({\bf p},x)\phi({\bf p},x)=const
\end{equation} where

\begin{equation}
\partial_{\tau}\Phi_{\tau}={\cal G}^{*}\Phi_{\tau}
\end{equation}
and ${\cal G}^{*}$ is the adjoint of ${\cal G}$ in $L^{2}(d{\bf
p}dx)$ (the Lebesgue measure $d{\bf p}$ is not Lorentz invariant,
the factor $\vert {\bf p}\vert^{-1}$ necessary for the Lorentz
invariance is contained in $\Phi$, see \cite{haba}). Hence, the
invariant measure $dxd{\bf p}\Phi_{R}$ is the solution of the
transport equation ${\cal G}^{*}\Phi_{R}=0$. Explicitly,
\begin{equation}\begin{array}{l}
\frac{1}{2}\gamma^{2}\partial^{j}\partial^{k}p_{j}p_{k}\Phi_{R}
-\frac{3}{2}\gamma^{2}\partial^{j}p_{j}\Phi_{R}-\gamma^{2}\partial^{j}R_{j}\Phi_{R}
=p_{\mu}\partial_{x}^{\mu}\Phi_{R}
\end{array}\end{equation}
(derivatives over the space-time coordinates will have an index
$x$). We denote by $\Phi_{ER}$ the $x$-independent solution of
eq.(16). Then, $R$ can be expressed in terms of $\Phi_{ER}$
\begin{equation}
R_{j}=\frac{1}{2}p_{j}+\frac{1}{2}p_{j}p_{k}\partial^{k}\ln
\Phi_{ER}
\end{equation}
We assume that $\Phi_{ER}$ is a function of the energy $p_{0}c$
multiplied by a constant  $\beta$ of the dimension inverse to the
dimension of the energy ($\beta=\frac{1}{kT}$ in the conventional
notation). In such a case  from eq.(17)
\begin{equation}R_{j}=\frac{1}{2}p_{j}+\frac{1}{2}\beta
c p_{j}\vert {\bf p}\vert (\ln\Phi_{ER})^{\prime}(c\beta\vert{\bf
p}\vert)
\end{equation}
Now, eq.(13) reads
\begin{equation}\begin{array}{l}
\partial_{\tau}\phi_{\tau}=\frac{1}{2}\gamma^{2}\triangle_{H}\phi_{\tau}
 +\frac{1}{2}\gamma^{2}p_{j}(1+\beta c\vert {\bf p}\vert
(\ln\Phi_{ER})^{\prime})\partial^{j}\phi_{\tau}
\end{array}\end{equation} and eq.(16)
\begin{equation}\begin{array}{l}
\frac{1}{2}\gamma^{2}\partial^{j}\partial^{k}p_{j}p_{k}\Phi_{R}
-2\gamma^{2}\partial^{j}p_{j}\Phi_{R}-\frac{1}{2}\gamma^{2}\beta
c\partial^{j} p_{j}\vert {\bf p}\vert (\ln\Phi_{ER})^{\prime}
\Phi_{R} =p_{\mu}\partial_{x}^{\mu}\Phi_{R}
\end{array}\end{equation}
Let us consider  an initial probability distribution of the form
$\Phi=\Phi_{ER}\Psi$. Then, from eq.(15) we obtain its evolution
\begin{equation}
\Phi_{\tau}=\Phi_{ER}\Psi_{\tau}
\end{equation}
as a solution of the equation (we do not make any  assumptions on
the form of $\Phi_{ER}$ here )
\begin{equation}
\begin{array}{l}
\partial_{\tau}\Psi_{\tau}=
\frac{1}{2}\gamma^{2}p_{j}p_{k}\partial^{j}\partial^{k}\Psi_{\tau}
+2\gamma^{2}p_{j}\partial^{j}\Psi_{\tau}+\frac{1}{2}\gamma^{2}p_{j}p_{k}(\partial^{j}
 \ln\Phi_{ER})
\partial^{k}\Psi_{\tau}-p_{\mu}\partial_{x}^{\mu}\Psi_{\tau}
\end{array}\end{equation}
\section{Spherical coordinates and the Kompaneetz equation}In this
section we restrict ourselves to the diffusion in momentum space.
The diffusion generator(7) is degenerate. This is easy to see if
we express it in the spherical coordinates on ${\cal H}_{0}$
\begin{equation}
p_{0}=r
\end{equation}
$ p_{1}=r\cos\phi\sin\theta $, $p_{2}=r\sin\phi\sin\theta
$,$p_{3}=r\cos\theta$. In these
coordinates\begin{equation}\begin{array}{l}
\triangle_{H}=r^{2}\frac{\partial^{2}}{\partial
r^{2}}+3r\frac{\partial}{\partial r}=\frac{\partial^{2}}{\partial
u^{2}}+2\frac{\partial}{\partial u}\end{array}
\end{equation}
where we  introduced an exponential parametrization of $r$
\begin{equation}
r=\exp u
\end{equation}
Here, $u$ varies over the whole real axis. It follows that the
three-dimensional diffusion of massive particles becomes
one-dimensional in the limit $m\rightarrow 0$.

In the coordinates (23) it is sufficient if we restrict ourselves
to the drags
\begin{equation}
X=\gamma^{2}R\frac{\partial}{\partial u} \end{equation} The
diffusion (12) is generated by
\begin{equation}
{\cal G}=\frac{\gamma^{2}}{2}(\partial_{u}^{2}+2\partial_{u})
+\gamma^{2}R\partial_{u}
\end{equation}

  We may write for integrals of spherically symmetric
functions
\begin{equation}
d{\bf p}=4\pi du\exp(3u) \end{equation} Then \begin{equation}
{\cal G}^{*}=e^{-3u}{\cal G}^{+}e^{3u} \end{equation}where
\begin{equation}\begin{array}{l}
{\cal G}^{+}=\frac{\gamma^{2}}{2}(\partial_{u}^{2}-2\partial_{u})
-\gamma^{2}\partial_{u}R
\end{array}\end{equation}
is the adjoint of ${\cal G}$ in $L^{2}(dudx)$. The probability
distribution evolves according to eq.(15). The invariant measure
$dxd{\bf p}\Phi_{R}\equiv dxdu\Phi_{I}$  solves an analog of
eq.(16)
\begin{equation} {\cal G}^{*}\Phi_{R}= {\cal
G}^{+}\Phi_{I}=0\end{equation}
where
\begin{equation}
\Phi_{I}=\exp(3u)\Phi_{R} \end{equation}

 We can express
eq.(31) as an evolution equation in the laboratory time $x^{0}$
\begin{equation}\begin{array}{l}
 \partial_{0}\Phi_{I}=\exp(-u){\bf p}\nabla_{{\bf x}} \Phi_{I}
\cr
+\frac{\gamma^{2}}{2}\exp(-u)(\partial_{u}^{2}-2\partial_{u})\Phi_{I}
-\gamma^{2}\exp(-u)\partial_{u}(R\Phi_{I})\end{array}
\end{equation}
where $ {\bf n}=\exp(-u){\bf p}$ is a unit vector in the direction
of motion (equations of this type describe a diffusive
perturbation of the propagation of a massless particle with the
velocity of light \cite{porous}).

The $x^{0}$ independent solution of eq.(33) is denoted
$\Phi_{EI}$. The drift $R$ is related to $\Phi_{EI}$
\begin{equation}
R=\frac{1}{2}(\partial_{u}\ln\Phi_{EI}-2)
\end{equation}If the diffusion (15) starts from an initial
distribution
\begin{equation}
\Phi_{I}=\Phi_{EI}\Psi^{I}
\end{equation}
then from eq.(15), using the equilibrium relation (34), we obtain
the evolution of $\Psi$
\begin{equation}\begin{array}{l}
\partial_{\tau}\Psi^{I}_{\tau}=\frac{1}{2}\gamma^{2}\partial_{u}^{2}\Psi^{I}_{\tau}
+
\frac{1}{2}\gamma^{2}(\partial_{u}\ln\Phi_{EI})\partial_{u}\Psi^{I}_{\tau}
\end{array}\end{equation}

 We consider J\"uttner equilibrium
distribution \cite{jut} in $L^{2}(du)$
\begin{equation}
\Phi_{EI}^{J}=r^{3}\exp(-\beta c r)\equiv r^{3}n_{J}
\end{equation}
Then, from eq.(34)
\begin{equation}
R=-\frac{1}{2}(-1+\beta c\exp (u))
\end{equation}The diffusion generator corresponding to the
J\"uttner distribution reads
\begin{equation}\begin{array}{l}
{\cal
G}=\frac{1}{2}\gamma^{2}\Big(\partial_{u}^{2}+3\partial_{u}-\beta
c\exp( u) \partial_{u}\Big)\cr =
\frac{1}{2}\gamma^{2}\Big(r^{2}\partial_{r}^{2}+4r\partial_{r}-\beta
cr^{2}
\partial_{r}\Big)\end{array}\end{equation}
For  the Bose-Einstein distribution (in Compton scattering the
photon number is preserved; if an equilibrium is achieved, then
the chemical potential $\mu\neq 0$, see \cite{wymann})
\begin{equation} \Phi_{EI}^{B}=r^{3}\Big(\exp(\beta (\mu
+cr))-1\Big)^{-1}\equiv r^{3}n_{E}(\mu)
\end{equation}
we have \begin{equation} R=-\frac{1}{2}\gamma^{2}\Big(-1+\beta
c\exp (u)\Big(1-\exp(-\beta\mu-\beta c\exp
u)\Big)^{-1}\Big)\end{equation} Eq.(36) for an
 evolution of $\Psi^{I}$ (with the drift (41) leading to the Bose-Einstein
equilibrium distribution) in the $r$-coordinates reads
\begin{equation}
\begin{array}{l}
\partial_{\tau}\Psi^{I}=\frac{1}{2}\gamma^{2}r^{-2}\Big(\partial_{r}r^{4}\partial_{r}\Psi^{I}
+r^{4}(\partial_{r}\ln
n_{E})\partial_{r}\Psi^{I}\Big)\end{array}\end{equation} where
\begin{equation}
\partial_{r}\ln n_{E}=-\beta c(1-\exp(-\beta\mu-\beta cr))^{-1}
\end{equation}
In the low temperature limit (or for the J\"uttner distribution
with the generator (39)) we have in the diffusion equation (42)
\begin{displaymath}
\partial_{r}\ln n_{E}\rightarrow -\beta c
\end{displaymath}

 We compare the relativistic diffusion equation with the
(non-linear) Kompaneetz equation usually written in the form
\cite{komp}
\begin{equation}
\partial_{\tau}
n=\kappa^{2}x^{-2}\partial_{x}\Big(x^{4}(\partial_{x}n
+n+n^{2})\Big)
\end{equation}
Eqs.(42) and (44) coincide if the non-gradient and friction
terms are neglected. Note that the form of the differential operators
on the rhs of eqs.(42) and (44) follows
from the conservation of probability for eq.(42) and conservation of
the number of photons for the Kompaneetz equation,
i.e.,\begin{displaymath}
\partial_{\tau}\int dr r^{2}\Phi_{\tau}=\partial_{\tau}\int dx x^{2}n_{\tau}=
0
\end{displaymath}
 In order to explore the appearance of friction
in eq.(44) we expand the photon distribution of the Kompaneetz
equation around its equilibrium value  in the same way as we
expanded the distribution of the diffusion process (15) around the
equilibrium in eq.(36). If we neglect $n^{2}$ on the rhs of
eq.(44) then the rhs disappears (because
$\partial_{x}n_{J}=-n_{J}$, with $x=\beta c r$) for the J\"uttner
distribution. Let\begin{displaymath}
n=\exp(-x)\Psi\end{displaymath} Then, neglecting the $n^{2}$ term
in the Kompaneetz equation (44) we obtain (after an elementary
rescaling $r\rightarrow \beta c r=x$,
$\kappa^{2}=\frac{1}{2}\gamma^{2}$) the diffusion equation (42)
with $n_{E}\rightarrow n_{J}$.

In general, the Kompaneetz distribution $n$ equilibrates to
$n_{E}$ (as $\partial_{x}n_{E}+n_{E}+n_{E}^{2}=0$ on the rhs of
eq.(44)). Let us write the initial distribution in the form
\begin{displaymath} n=n_{E}\Gamma
\end{displaymath}then the evolution of $\Gamma$ is determined by the
equation
\begin{equation}\begin{array}{l}
\partial_{\tau}\Gamma_{\tau}=\kappa^{2}x^{-2}\Big(\partial_{x}x^{4}\partial_{x}\Gamma_{\tau}
 +x^{4}(\partial_{x}\ln n_{E})\partial_{x}\Gamma_{\tau}\Big)
  +\kappa^{2}x^{-2}n_{E}^{-1}\partial_{x}\Big(x^{4}n_{E}^{2}(\Gamma_{\tau}^{2}-\Gamma_{\tau})\Big)\end{array}
\end{equation} The last term in eq.(45)  is absent in
eq.(42). It describes the effect of the quantum statistics which
promotes bosons to condense. As long as the density of photons is
low this  term will be negligible.

It remains to consider the Lorentz transformations of the
equilibrating diffusion equation. The diffusion equation (8)
(without friction) is explicitly Lorentz invariant as follows from
eq.(6). However, the equilibrium measure cannot be Lorentz
invariant if it is to be normalizable (in a finite volume) and if the
mass shell condition (1) is to be satisfied (there is only one
invariant measure on the mass-shell, this is $d{\bf
p}p_{0}^{-1}$). The equilibration to the J\"uttner or
Bose-Einstein equilibrium takes place in a preferred Lorentz frame
\cite{lorentz}. In a general frame described by the four-velocity
$w^{\nu}$ we should write the equilibrium distribution in the form
\begin{equation}
n_{E}(w,\mu)=\Big(\exp(\beta w^{\nu}p_{\nu}+\beta
w^{\nu}\mu_{\nu})-1\Big)^{-1}
\end{equation}
where (by convention) in the rest frame $w=(1,0,0,0)$ and
$\mu_{0}=\mu$.

We can write the Kompaneetz equation (44) in an arbitrary frame
\begin{equation}\begin{array}{l}
\partial_{\tau}n(w)=\frac{\kappa^{2}}{2}\Big(L_{\nu\rho}L^{\nu\rho}n(w)
\cr+2iw_{\nu}L^{\nu\rho}p_{\rho}n(w)(n(w)+1)+2w^{\nu}p_{\nu}n(w)(n(w)+1)\Big)
\end{array}\end{equation} In the rest frame eq.(47) coincides with eq.(44)
as can be checked using eq.(5) (with $L_{0j}=-i\vert {\bf p}\vert
\partial_{j}$ on the mass shell). In an arbitrary frame $n_{\tau}(w)$ has the
Bose-Einstein distribution $n_{E}(w,\mu)$ as an equilibrium
measure.

The Kompaneetz equation finds applications in astrophysics
\cite{rybicki}\cite{peebles}\cite{sun}

\cite{birk} \cite{stebb}\cite{bernstein}. The CMBR photons
described by $n_{E}(\mu=0)$ are scattered when passing clusters of
galaxies and other reservoir of hot plasma. Eq.(45) could be
solved with $n_{E}(\mu=0)$ and $\Gamma=1$ as an initial condition
(see the end of the next section). Then, it describes the photon
density evolution as it passes through the electron gas. The
resulting distortion of the black-body spectrum (Sunyaev-Zeldovich
effect, see \cite{birk}\cite{carl} for reviews)) can be compared
with observations (as discussed in \cite{habajpa} taking into
account the photon helicity does not change the diffusion
equations (8) and (13), the spin dependent part disappears in the
limit $m\rightarrow 0$). The Kompaneetz equation has been derived
from the Boltzmann equation \cite{komp}\cite{rybicki} under an
assumption that the electron velocities are non-relativistic.
Relativistic corrections to the Boltzmann equation describing the
Compton scattering in a gas of electrons and its diffusion limit
have been calculated in \cite{itoh}\cite{stebb}\cite{lesen}. These
computations involve relativistic corrections to the
photon-electron cross section  in the laboratory rest frame. They
are unrelated to the generalization (47) of the diffusion equation
to an arbitrary frame.

\section{Langevin stochastic equations}
We wish to express the solution of the diffusion equation (8)
 by a stochastic process
\begin{equation}
\phi_{\tau}(p,x)=E[\phi(p(\tau), x(\tau))]
\end{equation}
Here, \begin{equation}
x^{\mu}(\tau)=x^{\mu}+\int_{0}^{\tau}p^{\mu}(s)ds
\end{equation}
 The
 diffusion process  $p(\tau)$ as well as its generator ${\cal G}$
can be expressed in various coordinates. In the coordinates (7) we
have to find a frame $e_{ja}$ ($j,a =1,2,3$) such that
\begin{equation}
p_{j}p_{k}=e_{ja}e_{ka} \end{equation} A solution of eq.(50) is
\begin{equation} e_{ja}=p_{j}e_{a} \end{equation}where
$e_{a}e_{a}=1$. Then, the stochastic process solving eqs.(8) and
(48) reads \cite{ikeda}\cite{skorohod} \begin{equation}
dp_{j}=\frac{3}{2}\gamma^{2}p_{j}d\tau +\gamma p_{j}e_{a}db^{a}=
\gamma^{2}p_{j}d\tau +\gamma p_{j}e_{a}\circ db^{a} \end{equation}
where $b^{a}$ are independent Brownian motions defined as the
Gaussian process with the covariance
\begin{equation}
E[b^{a}(\tau)b^{c}(\tau^{\prime})]=\delta^{ac}min(\tau,\tau^{\prime})
\end{equation}
(we denote the expectation values by $E[...]$). The first of
eqs.(52) is the Ito equation whereas the circle in the second
equation denotes the Stratonovitch differential (the notation is
the same as in \cite{ikeda}). Eq.(52) is equivalent to the simpler
equation for $r=\vert {\bf p}\vert$
\begin{equation}
dr=\gamma^{2}rd\tau+\gamma r\circ db
\end{equation}
where $b$ is the one-dimensional Brownian motion. The solution of
eq.(52) is \begin{equation}
p_{j}(\tau)=p_{j}\exp(\gamma^{2}\tau+\gamma e_{a}b^{a}(\tau))
\end{equation}
 whereas the one of eq.(54)
 \begin{equation}r_{\tau}=r\exp(\gamma^{2}\tau+\gamma b(\tau))
\end{equation}
We could derive eq.(56) from eq.(55) setting $b=e_{a}b^{a}$.
Eq.(56) means that (in the coordinates (25))
\begin{equation}
u_{\tau}=u+\gamma^{2}\tau+\gamma b(\tau)
\end{equation}
When we know the probability distribution of $u_{\tau}$ then we
can calculate the probability distribution of the momenta. We have
(in the coordinates (23) the angles do not change in time)
\begin{displaymath}\begin{array}{l} E[f({\bf
p}_{\tau})]=E[f(r_{\tau},\phi,\theta)]=(2\pi\gamma^{2}\tau)^{-\frac{1}{2}}
\int_{0}^{\infty}
\frac{dy}{y}f(y,\phi,\theta)\exp\Big(-\frac{1}{2\tau\gamma^{2}}
(\ln\frac{y}{\vert {\bf p}\vert}-\gamma^{2}\tau )^{2}\Big)
\end{array}\end{displaymath}Hence, $\vert{\bf p}\vert $ has the log-normal distribution.
Note that if we wish to express the probability distribution
$\Psi$ (42) when $\partial_{r}\ln n_{E}\rightarrow 0$ in terms of
a stochastic process $\tilde{u}_{\tau}$ then $\tilde{u}_{\tau}$ is
the solution of the stochastic equation
\begin{displaymath}
d\tilde{u}_{\tau}=\frac{3}{2}\gamma^{2}d\tau+\gamma db_{\tau}
\end{displaymath}
In such a case
\begin{equation}\begin{array}{l}\Psi_{\tau} (r)= E[\Psi(\exp(\tilde{u}_{\tau}))]=(2\pi\gamma^{2}\tau)^{-\frac{1}{2}}
\int_{0}^{\infty}
\frac{dy}{y}\Psi(y)\exp\Big(-\frac{1}{2\tau\gamma^{2}}
(\ln\frac{y}{r}-\frac{3}{2}\gamma^{2}\tau )^{2}\Big)
\end{array}\end{equation}
is the solution of the equation
\begin{displaymath}
\begin{array}{l}
\partial_{\tau}\Psi_{\tau}=\frac{1}{2}\gamma^{2}r^{-2}\partial_{r}r^{4}\partial_{r}\Psi_{\tau}
\equiv {\cal G}^{*}\Psi_{\tau}\end{array}\end{displaymath} with
the initial condition $\Psi$. The solution (58) is discussed in
\cite{sun}\cite{bernstein}. For small $\tau$ and the initial
condition $\Psi=n_{E}(\mu=0)$ the solution can be approximated by
\begin{displaymath}
\Psi_{\tau}=n_{E}(\mu=0)+\tau{\cal G}^{*}n_{E}(\mu=0)
\end{displaymath}
This approximate solution of the diffusion equation without
friction is usually applied in a discussion of Sunyaev-Zeldovich
CMBR spectrum distortion
\cite{sun}\cite{birk}\cite{stebb}\cite{bernstein}\cite{carl}.

 The stochastic
equation for the diffusion with the J\"uttner friction (38)
generated by (39) reads
\begin{equation}
du=\gamma^{2} d\tau-\frac{1}{2}\gamma^{2}\beta
c\exp(u)d\tau + \gamma db_{\tau}
\end{equation}
It has the solution
\begin{equation}\begin{array}{l}
u_{\tau}=u+\gamma^{2}\tau+\gamma b_{\tau}-\ln\Big( 1
+\frac{1}{2}\gamma^{2}\beta c\exp(u)\int_{0}^{\tau} \exp(\gamma
b_{s}+\gamma^{2} s)ds\Big)\end{array}
\end{equation}
It is not easy to calculate the correlation functions of
$u_{\tau}$ or $r_{\tau}$ in a closed form (we could do it as an
expansion in $\beta$). However, correlation functions of
$\frac{1}{r_{\tau}}=\exp(-u_{\tau})$ can be calculated explicitly.
\section{A relation to the imaginary time quantum mechanics}

We can express solutions of the diffusion equations by an
imaginary time evolution generated by a quantum mechanical
Hamiltonian.The diffusion generator is of the form
\begin{equation}
{\cal G}=\frac{1}{2}\gamma^{2}\partial_{u}^{2}-\omega\partial_{u}
\end{equation}
Let
\begin{equation}
\Omega(u)=\int^{u}\omega
\end{equation}
Then
\begin{equation}
\exp(-\Omega){\cal G}\exp(\Omega)\equiv
-H=\frac{1}{2}\gamma^{2}\partial_{u}^{2}-V
\end{equation}
where
\begin{equation}
V=\frac{1}{2}\omega^{2}-\frac{1}{2}\partial_{u}\omega
\end{equation}
We have the standard Feynman-Kac formula
\cite{ikeda}\cite{skorohod} for $\exp(-\tau H)$
\begin{equation}
(\exp(-\tau H)\psi)(u)=E\Big[\exp\Big(-\int_{0}^{\tau}V(u+\gamma
b_{s})ds\Big)\psi(u+\gamma b_{\tau})\Big]
\end{equation}
Let
\begin{equation}
\psi=\exp(-\Omega)\phi
\end{equation}
Then, from eq.(63) it follows
\begin{equation}\begin{array}{l}
\phi_{\tau}(u)=(\exp(\tau {\cal
G})\phi)(u)=E[\phi(u_{\tau}(u))]=\exp(\Omega(u))(\exp(-\tau
H)\psi)(u)\end{array}
\end{equation}
where $u_{\tau}(u)$ (for the J\"uttner distribution) is the
solution (60) of the stochastic equation (59) with the initial
condition $u$. Inserting $\phi=1$ in eq.(67) we obtain $\exp(-\tau
H)\exp(-\Omega)=\exp(-\Omega)$. Hence,
\begin{equation}
H\exp(-\Omega)=0
\end{equation}
We can  relate statistical expectation values of the process
$u_{\tau}$ to some  expressions from an operator theory. Applying
eq.(67) we obtain
\begin{equation}\begin{array}{l}
\int du\exp(-\Omega(u))(\exp(-\tau H)\exp(-\Omega)\phi)(u)=
\int du \exp(-2\Omega(u))E[\phi(u_{\tau}(u))]\end{array}
\end{equation}
In the J\"uttner model
\begin{equation}
V=\frac{9}{8}\gamma^{4}+\frac{1}{8}\gamma^{4}\beta^{2}c^{2}\exp(2u)-
\frac{1}{4}(1+3\gamma^{2})\gamma^{2}\beta c\exp(u)
\end{equation}
The eigenstate of $H$ with the zero eigenvalue   is
\begin{equation}
\exp(-\Omega)=\exp\Big(\frac{3}{2} u-\frac{1}{2}\beta
c\exp(u)\Big)
\end{equation}
The statistical weight in eq.(69)
\begin{equation}
\exp(-2\Omega(u))=r^{3}\exp(-\beta c r)=\Phi^{J}_{EI}
\end{equation}
is just the J\"uttner distribution. It can be seen from eq.(69)
that in general \begin{displaymath} \exp(-2\Omega)=\Phi_{EI}
\end{displaymath}
for any equilibrium distribution $\Phi_{EI}$.

The potential   $V$ has a minimum. There is a well which supports
some bound states of $H$. Eq.(72) defines one of these
(normalizable) bound states. All the eigenstates of the potential
(70) are known ( this is the soluble Morse potential
\cite{flugge}) . On the set of functions
\begin{displaymath}
\phi(u)=\sum_{n}c_{n}\exp(-un)
\end{displaymath}
($n$ a natural number) the expectation value (67) could be
calculated explicitly using the solution (60). The formula
\begin{equation}
\exp(\tau{\cal
G})\phi=\sum_{k}a_{k}\exp(-\epsilon_{k}\tau)\phi_{k}
\end{equation}
allows to calculate the eigenvalues $\epsilon_{k}$ and
eigenfunctions $\phi_{k}$ from eqs.(60) and (67). We obtain in
this way the well-known eigenfunctions and eigenvalues of the
Morse potential \cite{flugge}. However, for an arbitrary friction
we would apply the relation (63) the other way round exploiting
quantum mechanical methods for estimates of the diffusion
equilibration.
\section{Discussion:The problem of time}
 In general, we have two candidates for time: $\tau$ and $x^{0}$.
In the relativistic dynamics, from the zero component of eq.(11),
we can express $\tau$ by $x^{0}$. Hence,  we can consider the evolution as a function of
$\tau$ or as a function of $x^{0}$. Let us discuss this problem
for the diffusion equation (22) assuming (for simplicity of the
argument) that $\Phi_{E}$ depends only on the momenta. First, we
find the solution $(x(\tau,x,{\bf p}),{\bf p}(\tau,{\bf p}))$ of
the stochastic equations
\begin{equation}\begin{array}{l}
dp_{j}=2\gamma^{2}p_{j}d\tau+\frac{1}{2}\gamma^{2}p_{j}p_{k}\partial^{j}\ln\Phi_{ER}d\tau
+\gamma p_{j}{\bf e}d{\bf b}(\tau)
\end{array}\end{equation}
\begin{equation}
dx^{j}=p_{j}d\tau
\end{equation}
\begin{equation}
dx^{0}=-\vert{\bf p}\vert d\tau\end{equation}
 with the initial condition $(x,{\bf p})$. Then,
\begin{equation}
\Psi_{\tau}(x,{\bf p})=E\Big[\Psi\Big(x(\tau,x,{\bf p}),{\bf
p}(\tau,{\bf p})\Big)\Big]
\end{equation}
is the solution of eq.(22).

We can obtain $\Phi_{R}$ from $\Phi_{\tau}$. For this purpose
integrate both sides of eq.(15) in the interval $[0,T]$. We have
\begin{equation} lim_{T\rightarrow\infty}T^{-1}\int_{0}^{T}d\tau \Phi_{\tau}=\Phi_{R}
\end{equation}
because the lhs of eq.(78) satisfies the equation ${\cal
G}^{*}\Phi_{R}=0$. Hence, if we denote \begin{equation}
\Phi_{\tau}=\Phi_{ER}\Psi_{\tau} \end{equation} Then, the
$\tau$-average of $\Psi_{\tau}$ satisfies the transport equation
(16)
\begin{equation}
\begin{array}{l}
\vert{\bf p}\vert \partial_{0}\Psi_{R}=
\frac{1}{2}\gamma^{2}p_{j}p_{k}\partial^{j}\partial^{k}\Psi_{R}
+2\gamma^{2}p_{j}\partial^{j}\Psi_{R}+\frac{1}{2}\gamma^{2}p_{j}p_{k}(\partial^{j}
 \ln\Phi_{ER})
\partial^{k}\Psi_{R}+p_{j}\partial_{j}\Psi_{R}
\end{array}\end{equation}
We can obtain a solution of eq.(80) from eq.(22) by a random
change of time. Let us treat the formula (49) for $x^{0}(\tau)$
 as a definition of $\tau$
\begin{equation}
\tau=\int_{0}^{x^{0}}\vert {\bf p}_{s}\vert^{-1}ds
\end{equation}
(here ${\bf p}_{s}$ is the solution of eq.(74); $x^{0}(\tau)$ can
be defined implicitly by (81)). We can see from eq.(81) that $\tau
$ depends only on events earlier than $x^{0}$. As a consequence
$p_{j}(x^{0})=p_{j}(x^{0}(\tau))$ is again a Markov process .
 Then, differentiating the momenta and coordinates according to the rules of the Ito calculus
 \cite{ikeda} we obtain the following Langevin equations (for
mathematical details of a random change of time see
\cite{skorohod}\cite{ikeda})
\begin{equation}\begin{array}{l}
dp_{j}(x^{0})=2\gamma^{2}\vert {\bf
p}(x^{0})\vert^{-1}p_{j}dx^{0}+\frac{1}{2}\gamma^{2}\vert {\bf
p}(x^{0})\vert^{-1}p_{j}p_{k}\partial^{j}\ln\Phi_{ER}dx^{0}
\cr
+\gamma p_{j}\vert {\bf p}(x^{0})\vert^{-\frac{1}{2}}{\bf
e}d{\bf b}(x^{0})
\end{array}\end{equation}
\begin{equation}
dx^{j}=p_{j}\vert {\bf p}(x^{0})\vert^{-1}dx^{0}
\end{equation}
Let  $\Psi({\bf x},{\bf p})$ be an arbitrary function of ${\bf x}$
and ${\bf p}$ and $(x(x^{0},x,{\bf p}),{\bf p}(x^{0},{\bf p}))$
the solution of eqs.(82)-(83) with the initial condition $({\bf
x},{\bf p})$ then
\begin{equation} \Psi_{R}(x^{0},{\bf x},{\bf p})=E\Big[\Psi\Big({\bf
x}(x^{0},{\bf x},{\bf p}),{\bf p}(x^{0},{\bf x},{\bf p})\Big)\Big]
\end{equation}
 is the solution of the transport equation (80) with the initial condition
 $\Psi({\bf x},{\bf p})$. We can prove that  eq.(80) is satisfied by differentiation
 of eq.(84) applying  the rules of the Ito calculus (or the well-known
 relation between  Langevin equation and the diffusion equation).
Hence,
 in principle,
 having a solution $E[\Psi({\bf x}_{\tau},{\bf p}_{\tau})]$ of the diffusion equation (22) as a function of $\tau $
 we can obtain a solution $E[\Psi({\bf x}(x^{0}),{\bf p}(x^{0}))]$
 of the transport equation (80)
 as a function of $x^{0}$. This is like in the deterministic case. The difference is in the
 averaging over events expressed by the expectation value (84)( a choice  of the time parameter and a random  change
 of time in a relativistic diffusion is also discussed in \cite{lasthang}).
Surprisingly, the new diffusion obtained in this way is the same
as the one derived by the averaging over $\tau$. If the limit
$\tau\rightarrow\infty$ of $\Psi_{\tau}$ exists then this limit is
equal  to the one resulting from the averaging (78) (i.e.,
$\Psi_{\infty}=\Psi_{R}$). We can show this property  using the
equivalence to quantum mechanics of sec.6 .  We apply the
expansion (73) of $\Psi_{\tau}$ and perform the $\tau$ averaging
term by term. We can see that
\begin{displaymath}
\lim_{T\rightarrow \infty}T^{-1}\int_{0}^{T}
d\tau\exp(-\epsilon_{k}\tau)
\end{displaymath}
exists and is different from zero if and only if $\epsilon_{k}=0$.
Hence, the averaging over $\tau$ gives the same result  as the
limit $\tau \rightarrow \infty$ of $\Psi_{\tau}$. In this sense
for large $\tau$ the probability distribution $\Psi_{\tau}$
depends only on $x^{0}$.

If the initial distribution $\Phi$ depends on $x^{0}$ then we must
perform the limit $\tau\rightarrow \infty $ (or averaging) in
order to get rid off the additional time in $\Phi_{\tau}$.
However, if $\Phi$ does not depend on $x^{0}$ (what is a
reasonable choice of physical initial phase space distribution)
then apart from  the procedure discussed above which leads from
$\Phi_{\tau}({\bf x},{\bf p})$ to $\Phi_{R}(x^{0},{\bf x},{\bf
p})$ there is still the question whether $\Phi_{\tau}({\bf x},{\bf
p})$ could possibly have a physical meaning (the solution of  the
Kompaneetz equation is of this type). It satisfies eq.(22) instead
of the transport equation (80). The difference in the $\tau$ and
$x^{0}$ evolutions is in the $p_{0}^{-1}$ factor in the collision
term of the Boltzmann equation. In the massive case $p_{0}\simeq
mc$ in the non-relativistic approximation. Hence, the difference
between $\tau$ and $x^{0}$ may be insignificant. However, in the
massless case $p_{0}=\vert {\bf p}\vert$ can never be approximated
by a constant. For a small time $\tau\simeq \vert {\bf p}\vert
^{-1}x^{0}$. We can make this replacement in all solutions of
proper time equations (in particular in the Kompaneetz equation).
However, in general, we should make the random time change
discussed in this section. Nevertheless, for an estimation of the
Sunyaev-Zeldovich CMBR spectrum distortion
\cite{sun}\cite{bernstein} \cite{birk} this problem is not
relevant because the time $\tau$ is small and chosen as a physical
parameter (plasma penetration depth).  In such a case  the
spectral deformation is $n_{E}+\tau{\cal G}^{*}n_{E}$ (as
discussed at the end of sec.5)

It can be seen from our discussion that only a  small part of the
theory of relativistic Brownian motion is applied in astrophysics.
The behavior of the diffusion at large time (apart from its final
effect: the equilibrium) seems to have no observational
consequences. However, with the increasing sophistication of the
astrophysical observations the large time effects of Compton
scattering may give important information about the sources of the
the CMBR distortions. In this paper we have completely neglected
the effect of the gravity. The relativistic diffusion of massive
particles on a general gravitational background is discussed in
\cite{jan}\cite{deb3}\cite{deb4}\cite{bal}. We  elaborate
relativistic diffusion with a friction leading to equilibration in
a gravitational field for massive as well as massless particles in
a forthcoming paper \cite{habafut}. In such a framework the role
of strong gravity (resulting from compact stars or dark matter)
and the effect of the expansion of the Universe upon the photon
diffusion can be investigated.

\end{document}